# Experimental Realization of a Reflective Optical Limiter

Jarrett H. Vella,[1] John H. Goldsmith,[1,2] Andrew T. Browning,[1,3] Nicholaos I. Limberopoulos,[1] Ilya Vitebskiy,[1] Eleana Makri,[4] and Tsampikos Kottos[4]

[1] The Air Force Research Laboratory, Sensors Directorate, Wright-Patterson Air Force Base, OH-45433, USA

[2] Wyle, Dayton, OH-45433, USA

[3] SelectTech Services Corporation, Centerville, OH-45433, USA

[4] Department of Physics, Wesleyan University, Middletown, CT-06459, USA

Optical limiters transmit low-intensity light, while blocking laser radiation with excessively high irradiance or fluence. A typical optical limiter involves a nonlinear material which is transparent at low light intensity and becomes opaque when the light intensity exceeds certain level. Most of the high-level radiation is absorbed by the nonlinear material causing irreversible damage. This fundamental problem could be solved if the state of the nonlinear material changed from transparent to highly reflective (not absorptive) when the intensity becomes too high. None of the known nonlinear optical materials display such a property. A solution can be provided by a nonlinear photonic structure. In this communication, we report the first experimental realization of a *reflective* optical limiter. The design is based on a planar microcavity composed of alternating $SiO_2$ and $Si_3N_4$ layers with a single GaAs defect layer in the middle. At low intensity, the planar microcavity displays a strong resonant transmission via a cavity mode. As the intensity increases, two-photon absorption in GaAs kicks in, initially resulting in the microcavity enhanced light absorption. Further increase in light intensity, though, suppresses the cavity mode along with the resonant transmission; the entire planar microcavity turns highly reflective within a broad frequency range covering the entire photonic band gap. This seemingly counterintuitive behavior is a general feature of resonant transmission via a cavity mode with purely nonlinear absorption.

## I. Introduction

Optical limiters are essential for protection of the human eye, optical sensors, and other optical and electronic devices from high intensity laser radiation. The existing passive optical limiters utilize nonlinear optical materials, which turn opaque when the irradiance or fluence exceeds certain limiting threshold. Such nonlinearity can be caused by two-photon absorption [1], reverse saturable absorption [2, 3], photo-conductivity, and other physical mechanisms (see for example [4 - 10] and references therein). A common problem with existing passive optical limiters is that the nonlinear optical material is directly exposed to the high level radiation, which can result in overheating, dielectric breakdown, or other irreversible damage. One possible solution is provided by photonic structures, such as a planar microcavity or a Fabry-Perot resonator, incorporating a low-loss nonlinear optical material with the refractive index dependent on the light intensity [11-20]. The limiting action in these cases can be achieved by a nonlinear shift of the transmission window of the photonic structure. Such a shift is inherently small and thus cannot provide broadband protection from high-power laser radiation. More importantly, the nonlinear material in such schemes is directly exposed to the high-level laser radiation, which can cause irreversible damage.

To address the above problems, a new concept of a reflective photonic limiter was introduced recently in [21, 22]. The proposed conceptual design, see Fig. 1, is based on the well-known phenomenon of resonant transmission via a nonlinear localized defect mode (hereinafter, a cavity mode). There are two key physical requirements to the constitutive optical materials of the planar microcavity in Fig. 1: (a) the defect layer must display purely nonlinear absorption – the linear absorption should be negligible; (b) the Bragg reflectors on both sides of the nonlinear defect layer must be linear with negligible losses, and they must have a much higher laser-induced damage threshold compared to the nonlinear material of the defect layer. At low intensity, such a planar microcavity displays a narrowband resonant transmission in the vicinity of the cavity mode. When the irradiance or fluence exceed certain level, nonlinear absorption suppresses the cavity mode, along with the resonant transmission; the entire layered structure acts as a Bragg mirror, reflecting most of the incident radiation back to space. Importantly, at the high reflectivity regime, only a tiny portion of the high-level input radiation reaches the defect layer, which protects the nonlinear optical material from laser-induced damage. Herein, the first experimental demonstration of such a planar microcavity is presented and evidence is provided that shows it indeed acts as a reflective optical limiter.

The paper is organized as follows. In Section II, the conceptual design of a reflective optical limiter is reviewed with emphasis on the physical requirements for the constitutive optical materials of the planar microcavity. In Section III, fabrication and testing procedures of the optical limiter are described. Section IV reports experimental results showing that the nonlinear planar microcavity does act as a reflective optical limiter.

## II. A specific realization of the concept of reflective photonic limiter

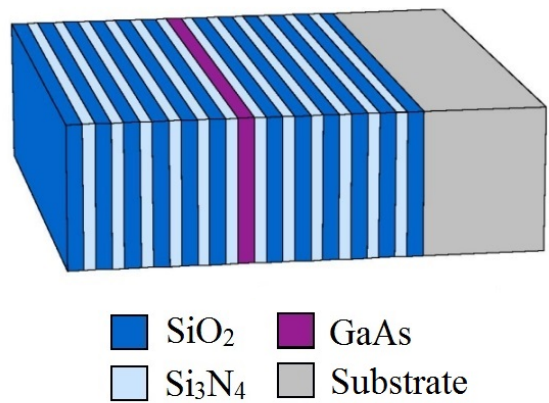

FIG. 1. Planar microcavity consisting of a half-wave GaAs layer (d=232 nm) sandwiched between two identical Bragg mirrors, each composed of six alternating quarter-wave layers of $SiO_2$ (d=264.8 nm) and $Si_3N_4$ (d=194.8 nm). The grey layer is the borosilicate glass substrate.

The planar microcavity shown in Fig. 1 presents a specific realization of a reflective optical limiter introduced, on a conceptual level, in Refs. [21, 22]. It is composed of a nonlinear layer sandwiched between two identical Bragg reflectors. The nonlinear (defect) layer is amorphous gallium arsenide (GaAs). The Bragg reflectors are composed of alternating silicon dioxide (SiO$_2$) and silicon nitride (Si$_3$N$_4$) quarter wave layers. In the near-infrared region, both silicon dioxide and silicon nitride are linear with negligible losses. Gallium arsenide is also virtually lossless, but only under low irradiance. As the value of light intensity $|E(z)|^2$ at the nonlinear layer location increases, the gallium arsenide displays a strong nonlinear (two-photon) absorption. Importantly, the field intensity $|E|^2$ in the vicinity of the defect layer (inside the microcavity) can be very different from that of the incident light. Specifically, at low irradiance, the value of $|E|^2$ at the defect layer is expected to be much higher than that of the incident light, while at high irradiance, the opposite is true, as depicted later in Fig. 3.

If the field intensity $|E(z)|^2$ at the nonlinear layer location is small enough, the imaginary part $\varepsilon''$ of the permittivity of the GaAs is negligible, and the planar microcavity supports a cavity mode with the frequency lying in the middle of the photonic band gap of the Bragg reflector. In the spectral vicinity of the cavity mode frequency, the planar microcavity in Fig. 1 displays a strong resonant transmission, shown in Fig. 2. The photonic bandgap of the Bragg reflector is clearly visible between 1385 nm and 1970 nm, while the sharp peak at 1633 nm (FWHM, 13.3 nm, 52% transmission) corresponds to the cavity mode resulting from incorporation of a single GaAs defect layer in between the two Bragg reflectors. Fringes on either side of the photonic bandgap are Fabry-Perot resonances, and for wavelengths shorter than 1385 nm, are attenuated by linear absorption from GaAs. The fact that the experimentally observed resonant transmissivity is below unity (0.52) can be attributed to structural imperfections and, in part, to a small but finite linear absorption of the GaAs layer; the latter is greatly enhanced by the planar microcavity, as depicted next in Fig. 3a. In our simulations, linear absorption and possible structural imperfections are neglected.

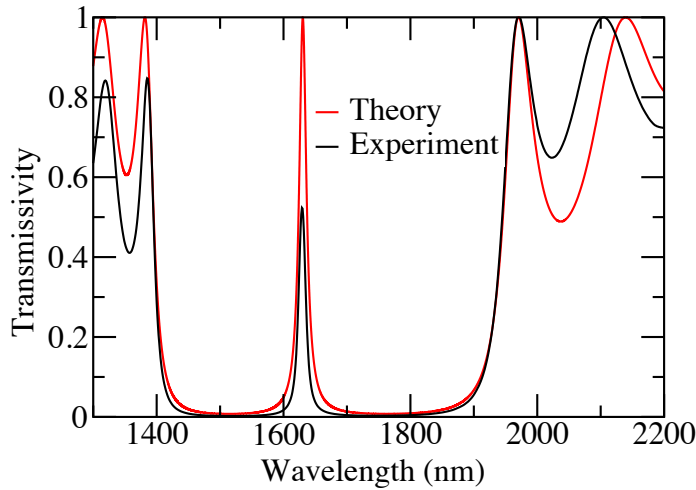

FIG. 2. Experimental (the black curve) and simulated (the red curve) low-intensity transmission spectrum of the planar microcavity in Fig. 1. The refractive indices used in the simulation are as follows: n (SiO$_2$) = 1.49, n (Si$_3$N$_4$) = 2.16, n (GaAs) = 3.42. The peak in the middle of the photonic band gap corresponds to the resonant transmission. If the incident light intensity increases, the cavity

mode will disappear and the layered structure will become highly reflective within the entire photonic band gap.

If the irradiance is increased, so is the field intensity $|E(z)|^2$ at the defect layer location. As a consequence, both the real and imaginary parts of the permittivity $\varepsilon = \varepsilon' + i\varepsilon''$ of the GaAs will change. The change in the real part, $\varepsilon'$, is small (less than 1% in our case). For shorter wavelengths, the nonlinear shift in $\varepsilon'$ could be much larger (see, for example, [23] and references therein), but not in our case where the photon energy is smaller than the GaAs band gap. The small shift in $\varepsilon'$ can only result in a small shift of the transmission window. If the transmission window shift is smaller than the transmission window itself and/or smaller than the laser pulse bandwidth, the shift itself does not provide protection from high-level laser radiation. In contrast, the increase in the imaginary part $\varepsilon''$ of the GaAs permittivity due to nonlinear absorption can be significant enough to completely suppress the cavity mode and render the entire structure in Fig. 1 highly reflective at the entire photonic band gap. To understand such a behavior, consider the simulated field distribution inside the planar microcavity of Fig. 1 at the frequency of transmission resonance. The computations were performed using the same standard transfer matrix formalism as in Refs. [21, 22] and presented in Fig. 3.

If the irradiance is low, the GaAs layer displays negligible absorption ($\varepsilon'' = 0$), and the steady-state field distribution inside the layered structure will look like that shown in Fig. 3a. In this case, the field intensity in the vicinity of the defect layer is much higher than that of the incident wave. This will result in the enhancement of nonlinear interactions and, eventually, will lead to a decrease of the limiting threshold, compared to that provided by a stand-alone nonlinear layer (see, for example, [24] and references therein).

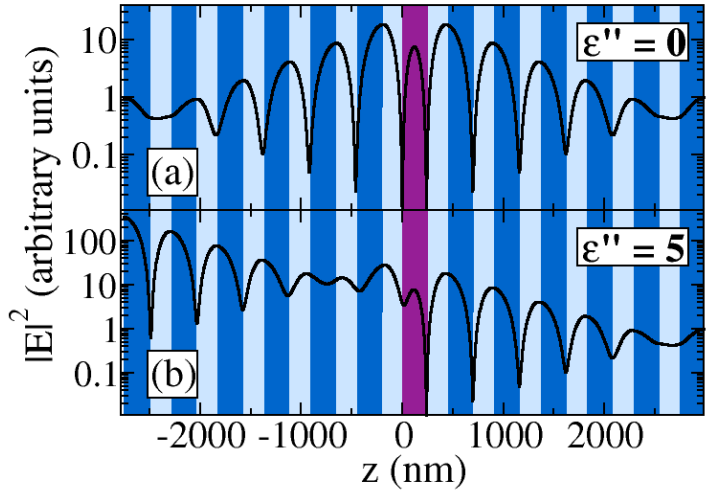

FIG. 3: Simulated field distribution $|E(z)|^2$ (solid black line) at the frequency of transmission resonance: (a) At low irradiance, the GaAs layer is virtually lossless and supports a cavity mode, providing strong resonant transmission shown in Fig. 2. The field intensity in the vicinity of the defect layer is much higher than that of the incident wave. (b) Under high irradiance, the increased

(nonlinear) losses in the GaAs layer suppress the cavity mode, along with the resonant transmission. The field intensity $|E(z)|^2$ in the vicinity of the defect layer is now much lower than that of the incident wave. The latter amounts to shielding the nonlinear layer from high-level laser radiation.

If the irradiance is well above the limiting threshold, the nonlinear GaAs layer becomes quite lossy, as seen in Fig. 4. We can speculate that the governing physical mechanism responsible for the nonlinear losses in amorphous GaAs is two-photon absorption, the same as in single-crystalline GaAs, but we know too little about amorphous GaAs to try to describe quantitatively the nonlinear absorption in it and to solve numerically the nonlinear scattering problem for the layered structure in Fig. 1. Instead, we invoke the following important qualitative conclusion drawn in [21, 22]. We assume that, regardless of the specific physical nature of nonlinear absorption in the defect layer, as soon as the nonlinear losses reach a certain level, the cavity mode becomes suppressed along with the resonant transmission, leading to high reflectance of the entire stack. The simplest way to qualitatively describe the effect of nonlinear absorption in the defect layer is to treat the scattering problem linearly, while assuming that the effective value of $\varepsilon''$ in the nonlinear layer is a growing function of the field intensity $|E(z)|^2$ at the defect layer location. The effective value of $\varepsilon''$ might also depend on the pulse shape and duration, which in our experiment are fixed.

The results of such simulations for a particular value of the effective $\varepsilon''$ are presented in Fig. 3b. The field distribution in Fig. 3b clearly shows that the defect (cavity) mode is suppressed, and the field intensity $|E(z)|^2$ in the vicinity of the nonlinear defect layer is much lower than that of the incident wave. The latter amounts to shielding of the vulnerable nonlinear layer from direct exposure to the high-level laser radiation. It also implies a dramatic increase in the limiter damage threshold, compared to the case of a stand-alone nonlinear layer. Such a shielding effect takes place when the imaginary part $n''$ of the effective refractive index of the nonlinear layer is of the order of unity. In the example shown in Fig. 3b we used $\varepsilon'' = 5$, but the resonant field distribution inside the layer structure remains qualitatively the same for $\varepsilon''$ values anywhere between, for example, 1 and 10. Upon inspection of Fig. 4, those values are qualitatively consistent with the experimentally observed nonlinear absorption in amorphous GaAs.

The dynamic range of a limiter is usually defined as the ratio of the limiter damage threshold and the limiting threshold. The above semi-qualitative consideration shows that the dynamic range of the planar microcavity in Fig. 1 is much higher than that of a stand-alone GaAs layer due to the simultaneous increase in the damage threshold and reduction in the limiting threshold.

The conditions under which the photonic structure turns reflective depend on: (i) the choice of nonlinear material of the defect layer, (ii) the number of bilayers in the Bragg mirrors in Fig. 1, and (iii) the laser pulse shape and duration. For instance, the multilayer can become highly reflective if the peak irradiance exceeds certain level [13]. Alternatively, the planar microcavity can become reflective when the pulse fluence exceeds certain level [14]. In the former case, we have a reflective irradiance limiter, while the latter case corresponds to a reflective fluence limiter. In many cases, though, both the pulse duration and the peak irradiance are equally important. In practical terms, it implies the possibility of simultaneous protection from laser pulses with excessively high peak irradiance and/or excessively high fluence.

Let us reiterate that a planar microcavity similar to that shown in Fig. 1 will act as a reflective optical limiter if the following physical requirements are satisfied:

1. The constitutive materials of the Bragg mirrors ($SiO_2$ and $Si_3N_4$ in Fig. 1) have negligible losses at the frequency range of interest, and their laser-induced damage threshold is high.
2. The defect layer (GaAs in Fig. 1) displays a significant nonlinear absorption (such as two-photon absorption), but negligible linear absorption at the wavelength of interest.

If the first of the two conditions is violated, the structure in Fig. 1 would still act as an optical limiter, but it would be an absorptive limiter – not a reflective one. A violation of the second condition would result in suppression of the low-level resonant transmission, making the stack act as a Bragg reflector regardless of the incident light irradiance/fluence. As long as the above two conditions are satisfied, one can always adjust the layer thicknesses so that the low-intensity resonant transmission occurs at a desired frequency in the middle of a photonic band gap. The irradiance or fluence above which the optical limiter becomes reflective can be adjusted within a wide range by changing the number of periods (bilayers) in the Bragg mirror in Fig. 1.

The specific choice of optical materials presented in Fig. 1 satisfies the above conditions. For wavelengths longer than 800 nm, pure, crystalline GaAs has negligible absorption [25]. For the amorphous GaAs used in this study, linear absorption of a standalone film was not observed in the vicinity of the cavity mode.

## III. Experimental Details

Plasma enhanced chemical vapor deposition (PECVD) was used to deposit the Bragg reflector onto amorphous borosilicate glass. The Bragg reflector consisted of a 264.8 nm thick $SiO_2$ layer, on top of which a 194.8 nm thick layer of silicon nitride $Si_3N_4$ was deposited. This bilayer structure was deposited six successive times in one PECVD run. Following deposition of the GaAs defect layer, the order of the Bragg reflector layers was reversed; silicon nitride was deposited first, followed by silicon dioxide. According to [26], the refractive indices for silicon dioxide and silicon nitride are $n(SiO_2) = 1.49$ and $n(Si_3N_4) = 2.16$, respectively. These values are consistent with the photonic band-gap location in Fig. 2. The refractive index of the amorphous GaAs can be extracted from the cavity mode location in Fig. 2. The respective value $n(GaAs) = 3.42$ is consistent with that from [27].

Gallium arsenide was deposited at room temperature by using a GCA RF diode sputtering system. The chamber was evacuated to a base pressure less than $1 \times 10^{-6}$ Torr and then backfilled with 10 mTorr of argon. After cleaning the target for 5 minutes, sputtering was commenced by applying a bias of 175W RF to a 5" GaAs target placed 4 cm above the substrate with the Bragg reflector. X-ray 2theta-omega diffraction patterns were obtained for the subsequent film using a PANalytical Empyrean x-ray diffractometer equipped with a PIXcel scanning line detector. The 2theta-omega x-ray diffraction patterns lacked the presence of any peaks indicating that the GaAs film was amorphous.

Next, we proceed with the nonlinear characterization. Steady state transmission spectra were acquired using a Cary 5000 UV-VIS-NIR absorption spectrophotometer. Reflective optical limiting behavior was characterized with the I-scan method [28]. A Spectra-Physics Solstice Ti:$Al_2O_3$ laser with a Gaussian, 150 fs pulsewidth (measured at the sample) and 1 kHz repetition rate was used to pump a TOPAS optical parametric amplifier. The laser power was measured using an Ophir 3A-FS thermopile. Attenuation of the incident power was achieved using a pair of crossed, linear polarizers. The laser was focused onto the sample using a 100 mm focal length lens. Light was incident at an angle of 6$^o$ to enable the collection of reflection spectra.

Transmitted or reflected light was focused onto the entrance slit of a Horiba-Jobin Yvon iHR320 spectrometer, equipped with a 2500 nm blaze, 120 g/mm diffraction grating and an extended InGaAs low noise, single channel detector connected to an SR830 lock-in amplifier. Four reflective 1.0 absorbance neutral density filters were positioned in front of the entrance slit to prevent detector saturation. A 200 mm focal length, on-axis spherical mirror was used to collect reflected light and focus it onto the spectrometer entrance slit. Power dependent reflectivity and transmissivity measurements are obtained from spectral data. The full transmission and reflection spectrum for each incident power level was obtained for both the reflective optical limiter and the free laser pulse. Reflection spectra of the laser pulse were obtained by substituting a gold mirror for the Bragg reflector. In this way, accurate transmissivity and reflectivity values were obtained, regardless of any wavelength shifts associated with nonlinear absorption in the defect layer.

*IV. Results and Discussion*

The as-deposited silicon dioxide, silicon nitride, and gallium arsenide films were all amorphous, as determined by x-ray diffraction. Gallium arsenide is a direct bandgap semiconductor, with two-photon absorption in the near-infrared to mid-infrared regions [29,30]. However, the GaAs typically studied for nonlinear measurements is crystalline. Because the underlying silicon nitride layer is amorphous, annealing the GaAs into a crystalline material would be difficult, if not impossible. Lederer and co-workers [31] studied the nonlinear absorption of GaAs optical limiters and demonstrated that although the efficacy is reduced, GaAs can still be an effective two-photon absorption material, even when amorphous. To verify that the as-deposited amorphous GaAs is capable of two-photon absorption, irradiance dependent absorption, transmission, and reflection measurements were carried out with a laser pulse centered at 1600 nm (full width at half-maximum, FWHM, 40.5 nm) for a standalone GaAs film of similar thickness, as illustrated in Fig. 4. These measurements correspond to the spectrum of the laser pulse. At low peak irradiance (< 1 GW/cm$^2$), the sum of the transmitted and reflected energy is approximately equal to the incident energy, indicating that for this film thickness, the linear absorption is negligible. As the peak irradiance increases to 50 GW/cm$^2$, the nonlinear absorption increases, indicating the as-deposited, amorphous GaAs does display two-photon absorption at the wavelengths of interest.

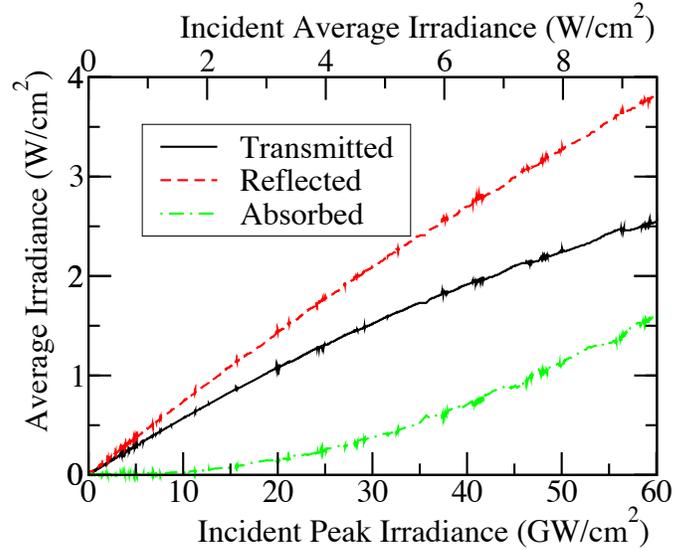

FIG. 4. Irradiance dependent transmission (solid black line), reflection (dashed red line), and absorption (dash-dotted green line) for a neat, amorphous gallium arsenide film 232 nm thick on glass, for a laser pulse centered at 1600 nm (FWHM 40.5 nm).

Representative irradiance-dependent, normalized transmissivity spectra for the planar microcavity are presented in Fig. 5. For an average irradiance of 0.82 W/cm$^2$ (5.18 GW/cm$^2$ peak irradiance, black line) and 7.39 W/cm$^2$ (46.3 GW/cm$^2$ peak irradiance, red line), the transmissivity decreases from a maximum of 0.42 to 0.024, respectively. The FWHM spectral widths are 12.45 nm and 15.71 nm, respectively. Detector noise prevents an accurate determination of the transmissivity maximum wavelength, so a wavelength range of 1630-1636 nm is reported. The accuracy of the spectrometer used to collect Fig. 5 data is ± 1 nm; the 0.82 W/cm$^2$ transmissivity spectral width is statistically equal to the spectral width of the steady-state transmission spectrum in Fig. 2. As the average irradiance increases to 7.39 W/cm$^2$, the spectral width increases to 15.7 nm. This slight broadening of the cavity mode transmission band is a result of modulations in the refractive index caused by two photon absorption as described in Section II.

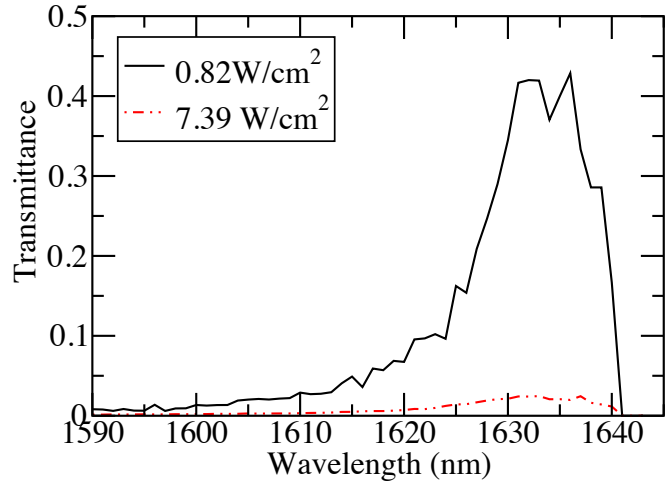

FIG. 5. Transmittance spectra for some representative values of the average irradiance: 0.821 W/cm$^2$ (5.18 GW/cm$^2$ peak irradiance, black line) and 7.39 W/cm$^2$ (46.3 GW/cm$^2$ peak irradiance, red line). The structure is a planar microcavity consisting of a half-wave GaAs layer (d=232 nm) sandwiched between two identical Bragg mirrors, each composed of six alternating quarter-wave layers of SiO$_2$ (d=264.8 nm) and Si$_3$N$_4$ (d=194.8 nm).

Fig. 6 depicts representative nonlinear reflection spectra for an average irradiance of 1.08 W/cm$^2$ (6.77 GW/cm$^2$ peak irradiance) and 6.02 W/cm$^2$ (37.7 GW/cm$^2$ peak irradiance). Under low irradiance, the reflectivity takes the minimal value of 0.31. The exact reflectivity maximum wavelength cannot be elucidated from the spectrum, but it is between 1630 nm-1639 nm. At higher average irradiance, 6.02 W/cm$^2$, the reflectivity spectrum is constant, between 0.9-1.0 for all frequencies inside the bandgap. It is clear from Fig. 6 that some transmission is present, but detector noise prevents the expected reflection decrease from being observed. The transition from low to almost total reflectivity indicates that the layered structure in Fig. 1 does act as a reflective optical limiter.

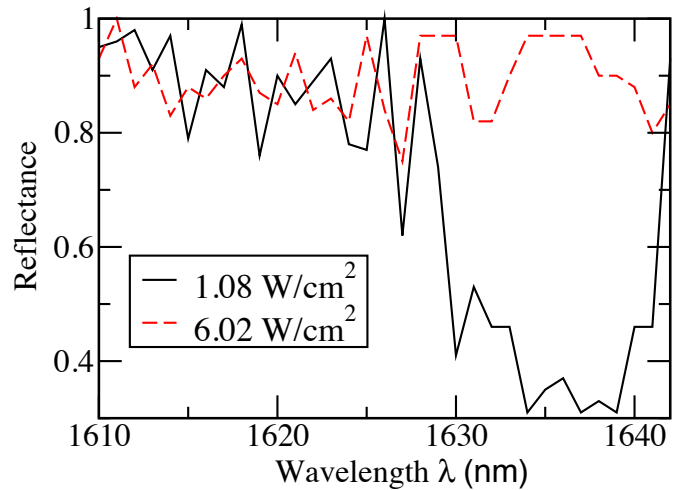

FIG. 6. Representative nonlinear reflectance spectra for an average irradiance of 1.08 W/cm$^2$ (6.77 GW/cm$^2$ peak irradiance, black line) and 6.02 W/cm$^2$ (37.7 GW/cm$^2$ peak irradiance, red line).

The full set of irradiance dependent reflectivity and transmissivity can be found in Fig. 7. Peak irradiance values ranging from 4.24 GW/cm$^2$ (0.827 W/cm$^2$ average irradiance) to 46.3 GW/cm$^2$ (7.39 W/cm$^2$ average irradiance) were studied. Near the resonant transmission maxima, the transmissivity decreases from 0.41 at 4.24 GW/cm$^2$ to 0.026 at 46.3 GW/cm$^2$. The reflectivity shows a concomitant increase from 0.24 at the lowest irradiance to greater than 0.98 at the highest irradiance. At the higher irradiance levels, detector noise becomes a larger fraction of the spectral transmission intensity. In addition, instrumental limitations prevented us from the simultaneous measurement of transmission and reflection spectra under the same irradiances. For these reasons, the sum of the reflectivity and transmissivity appears artificially higher than unity for certain levels of irradiance.

It is very clear that the optical limiting ability of the reflective limiter is superior to a standalone GaAs thin film. Taking an average irradiance of 4.00 W/cm$^2$ (25.0 GW/cm$^2$ peak) as the point where the limiter has "turned on", the extracted reflectivity is 0.97, while the extracted transmissivity is 0.15. According to Fig. 4, at the same average irradiance, the transmissivity and reflectivity of the standalone GaAs film are 39% and 54%, respectively. Apart from the fact that the stand-alone GaAs film does not provide wavelength discrimination, we can see that the Bragg reflectors in Fig. 1 greatly enhance the limiting properties of GaAs film.

Our experimental observations strongly indicate that the structure behaves as a *reflective* optical limiter, in agreement with the predictions of Refs. [21, 22]. Upon visual inspection, no optical damage could be observed for the reflective optical limiter after testing, indicating it can provide continuous, uninterrupted protection from high level radiation.

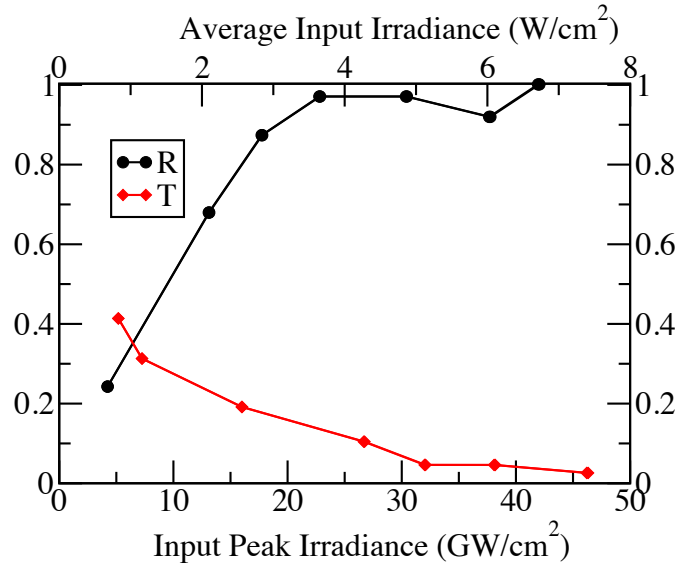

FIG. 7. Transmissivity and reflectivity versus the peak irradiance (the bottom horizontal axis) and the average irradiance (the top horizontal axis). The laser pulse duration is 150 fs with the repetition rate of 1 kHz. The laser frequency is adjusted to coincide with the maximum transmission at the respective incident light intensity. Under high irradiance, the layered structure becomes highly reflective within frequency range covering the entire photonic band gap. Such a behavior is characteristic of a reflective optical limiter. The structure is a planar microcavity consisting of a half-wave GaAs layer

(d=232nm) sandwiched between two identical Bragg mirrors, each composed of six alternating quarter-wave layers of $SiO_2$ (d=264.8 nm) and $Si_3N_4$ (d=194.8 nm).

## V. Conclusions

The first experimental demonstration of a reflective optical limiter is presented. The nonlinear planar microcavity consisting of a nonlinear GaAs layer placed between two Bragg mirrors, displays a narrowband low-intensity resonant transmission. In the case of femtosecond pulses with high peak irradiance, the resonant transmission disappears, and the planar microcavity turns highly reflective within a broad frequency range, covering the entire photonic band gap of the Bragg mirrors. By comparison, a stand-alone GaAs layer would also act as a nonlinear optical limiter at the same frequency range, but it would be an absorptive limiter – not a reflective one. The limiting threshold of the nonlinear microcavity in Fig. 1 is expected to be much lower, while its damage threshold – much higher than that of the stand-alone GaAs layer. The role of the layered structure is three-fold. Firstly, it makes the optical limiter reflective, rather than absorptive. Secondly, it shields the vulnerable nonlinear layer (GaAs) from laser-induced damage, thereby greatly increasing the dynamic range of the limiter. Finally, our planar microcavity design provides much stronger transmitted light attenuation above the limiting threshold. The above features can be attractive for various applications, including sensor protection and mode locking for the generation of ultrashort laser pulses. The current design can only perform at shortwave IR, where GaAs displays negligible linear absorption and very strong nonlinear two-photon absorption. With judicious choice of optical materials, the same principle can certainly be replicated for other wavelength ranges.

*Acknowledgement*

We acknowledge AFOSR support via LRIR09RY04COR and MURI No. FA9550-14-1-0037 (P.O. - Dr. Nachman).